\documentclass[namedreferences]{solarphysics}
\usepackage[optionalrh]{spr-sola-addons} % For Solar Physics
\usepackage{graphicx}                    % For eps figures, newer & more powerfull
\usepackage{color}                       % For color text: \color command
\usepackage{url}                         % For breaking URLs easily trough lines
\newcommand{\adv}{    {\it Adv. Space Res.}}

\newcommand{\aap}{    {\it Astron. Astrophys.}}

\newcommand{\apj}{    {\it Astrophys. J.}}
\newcommand{\apjl}{   {\it Astrophys. J. Lett.}}

\newcommand{\grl}{    {\it Geophys. Res. Lett.}}

\newcommand{\jgr}{    {\it J. Geophys. Res.}}

\newcommand{\solphys}{{\it Solar Phys.}}

\newcommand{\ssr}{    {\it Space Sci. Rev.}}

\begin{document}

\begin{article}

\begin{opening}

\title{On existence of two different mechanisms for forming coronal mass ejections}

%%%%%%%%%%%%%%%%%%%%%%%%%%%%%%%%%%%%%%%%%%%%%%%%%%%
%% Authors Names
%
\author{V.~\surname{Eselevich}$^{1}$\sep
        M.~\surname{Eselevich}$^{1}$\sep
        V.~\surname{Romanov}$^{2}$\sep
        D.~\surname{Romanov}$^{2}$\sep
        K.~\surname{Romanov}$^{2}$
              }

%%%%%%%%%%%%%%%%%%%%%%%%%%%%%%%%%%%%%%%%%%%%%%%%%%%
%% Runningheads
%
\runningauthor{Eselevich et al.}

\runningtitle{On existence of two different mechanisms}

%%%%%%%%%%%%%%%%%%%%%%%%%%%%%%%%%%%%%%%%%%%%%%%%%%%
%% Affilations
%
\institute{$^{1}$ Institute of solar-terrestrial physics, Irkutsk,
Russia \\ email: \url{esel@iszf.irk.ru}}

\institute{$^{2}$ Krasnoyarsk state pedagogical university,
Krasnoyarsk, Russia}

%%%%%%%%%%%%%%%%%%%%%%%%%%%%%%%%%%%%%%%%%%%%%%%%%%%
%%% Abstract
\begin{abstract}
We confirm the principal difference of the initiation phase
between the impulsive and gradual CME motion trajectory revealed
earlier in preliminary studies. Based on studying the dynamics of
two impulsive CME (25 March 2008 and 13 June 2010), and also the
MHD-approximation computations, we have come to a conclusion that
forming impulsive CME starts under the solar photosphere and may
be associated with supersonic emergence of magnetic tubes from the
convective region. A radial velocity of such tubes at the
photosphere level can reach hundreds of km~s$^{-1}$, and their
angular size $d\approx$ (1-3)$^\circ$. A probable reason of their
rise from the convective region is the ``slow wave'' instability
(the Parker instability).
\end{abstract}

%%%%%%%%%%%%%%%%%%%%%%%%%%%%%%%%%%%%%%%%%%%%%%%%%%%
%% Keywords
%

\end{opening}
%-------------------------------------------------

%%%%%%%%%%%%%%%%%%%%%%%%%%%%%%%%%%%%%%%%%%%%%%%%%%%
%% Sections
%
\section{Introduction}

A mechanism for forming coronal mass ejections (CME) has been the
main problem in their studying over several decades. For years, a
solar flare has been considered as a CME's possible source.
Theoretical papers consider the reconnection of magnetic field
opposite polarity lines as the only mechanism for a flare
\cite{Giovanelli1946,Priest1982,Priest2000,Somov2007}. MHD theory
predicts that the energy released during this process is shared
roughly equally between plasma heating and kinetic energy of
plasma ejection -- ``flare CME'' or ``flare jet''
\cite{Priest2000}. Direct observations confirm the presence of the
reconnection mechanism during flares \cite{Lin2005}. But the
released energy distribution between various plasma components
differs significantly from the MHD prediction. Thus, in
\inlinecite{Sui2005} it was experimentally shown that in a
mean-power isolated flare, with an X-ray class $\geq M1$, the
flare's total energy of $E_h\sim (2-3)10^{30}$ erg was distributed
approximately as follows:
\begin{itemize}
\item energy of accelerated particles (electrons) $E_p/E_h\approx
(70-75)\%$;
\item thermal energy of plasma $E_T/E_h\approx (20 -
25)\%$;
\item kinetic energy of the plasma flow from the Sun
(``flare CME (jet)'')\\ $E_{cme}/E_h\sim 1\%$.
\end{itemize}
Taking into account the error associated with an inaccuracy in
volume the heated plasma and the accelerated electrons energy
spectrum lower boundary \cite{Sui2005}, one may state the
following:
\begin{itemize}
\item energy of accelerated particles exceeds the plasma thermal
energy $E_T$ by a factor of 2-3;
\item energy of a ``flare CME (jet)''
makes few percent of the total released energy $E_h$.
\end{itemize}
The fact that during flares the energy bulk turns into
acceleration of charged particles (electrons) tells about an
essential role of the collective processes associated with the
turbulent oscillation excitation. For most powerful flares with an
X-ray class $\geq X1$, accelerated particles include, except
electrons, protons, ions of other elements and nuclei, and their
total energy exceeds $E_T$ by almost a factor of 4-5
\cite{Emslie2004}. Also, the share of a ``flare CME (jet)'' energy
may be under $0.01E_h$.

Thus, in all flares, regardless of their power, the reconnected
magnetic field energy majority falls on the collective particle
acceleration, and the minority falls on plasma heat, and still a
greater minority falls on kinetic energy of a ``flare CME (jet)''.
At the same time, the kinetic energy of the most observed powerful
CMEs that are flare-associated in their origin by time and
locality, often twice and more exceeds the total energy of
accelerated particles and released plasma heat \cite{Emslie2004}.
Hence, it follows that the powerful accompanying CMEs observed
simultaneously with flares are not ``flare CME (jet)'' and have
another origin \cite{Hundhausen1994}. They exceed ``flare CMEs
(jets)'' both in their size, and in their characteristic
velocities. Therefore, one can observe ``flare CME (jet)'' only
when there is no accompanying CME, i.e. in the events with
relatively small released energy. The reason for a ``flare CME
(jet)'' origin is the magnetic flux slowly emerging from under the
photosphere and its reconnection at its exit from the photosphere
with the surrounding magnetic field of opposite polarity. Such a
pattern has been studied well theoretically
\cite{Shimojo1996,Antiochos1998,Filippov1999,Asai2008,Masson2009}
and confirmed experimentally \cite{Liu2011}.

Powerful CMEs (``non-flare'') are classified into two groups by
their motion characteristics: ``gradual'' (slowly evolving) and
``impulsive'' \cite{Sheeley1999}. Impulsive are the fastest CMEs
that are accelerated near the solar surface at altitudes under
$0.2R_{\odot}$ relative to the limb \cite{MacQueen1983}. However,
fast CMEs may also origin as a result of a less fast acceleration
occurring within up to several solar radii
\cite{Plunkett2000,Yurchyshyn2002}.

In \inlinecite{Eselevich2011} it was shown that the parameters
reflecting the difference between the impulsive and gradual CMEs
in the physical nature of their origin are the CME's location,
velocity and angular size at their origin. The gradual CME origin
location is in the corona at $0.1R_\odot<h\leq 0.7R_\odot$ above
the solar limb. They start their motion having an angular size
within $\approx$ (15-65)$^{\circ}$ with an initial velocity
$V_0\approx 0$. Also, the flares that often accompany gradual CMEs
\cite{Hundhausen1994,Zhang2001} are not the main power source for
them, and reflect the action of the trigger mechanism accompanying
the CME motion start.

These peculiarities of the gradual CME origin and propagation
qualitatively and, in some characteristics, quantitatively agree
with the theory \cite{Chen1996,Krall2000} that considers an
eruption or sudden sun-outward motion of the magnetic flux rope
localized in the solar corona the source of gradual CMEs. In a
stationary state prior to eruption, the rope represents a
plasma-filled magnetic field screw line arc pattern whose two
footpoints are implanted in the solar photosphere. Along the rope,
a prominence substance may be placed. An eruption probable reason,
as per \inlinecite{Krall2000}, may be, for example, a fast
increase in the magnetic flux poloidal component in the rope.
Possible are also some other reasons \cite{Kuznetsov2000}. Also
considered is a reconnection mechanism in the vertical current
layer that is formed below the magnetic rope \cite{Vrsnak2004}. In
fact, various mechanisms may be in effect simultaneously. As per
\inlinecite{Forbes2000}, the onset of this or that mechanism
leading to the eruption of the resting rope in the corona may be
caused either by photospheric motions at the magnetic rope feet,
or by a new magnetic flux emerging from under the photosphere.

Models for slowly emerging (at subsonic velocity) magnetic rope
from a small depth from under the photosphere ($\approx -2000$ km)
and its successive eruption in the corona like a CME due to
various mechanisms have been considered in a number of papers
(e.g., \opencite{Archontis2004}; \opencite{Fan2005};
\opencite{Gibson2006}; \opencite{Archontis2010}). In these models,
the rope velocity is close to zero or very small prior to
eruption, i.e., they also describe gradual CMEs.

As per experimental studies \cite{Zhang2001}, the gradual CME
trajectory is characterized by three phases: the ``initial phase''
at which the rope velocity slowly grows from zero; the ``impulsive
phase'' of fast acceleration and the ``propagation phase'' with
approximately constant velocity.

Despite the relevance of the slowly emerging magnetic rope models,
they possess a serious deficiency: this is an arbitrary selection
of initial boundary conditions. Thereupon, of particular interest
are studies of the convective region magnetic tube dynamics
associated with the Parker instability
\cite{Romanov1993a,Fan1994}. The problem of the magnetic tube rise
from the convective region arbitrary depth is solved
self-consistently in them. The computations made in
\citeauthor{Romanov1993a}
(\citeyear{Romanov1993a},\citeyear{Romanov1993b}),
\inlinecite{Romanov2010} and having a preliminary character showed
a principal possibility of magnetic tube rise into the solar
atmosphere at high supersonic velocities. They revealed a
dependence of the rise resultant velocity on: a) a magnetic tube
emerge initial depth, b) an initial disturbance cross size, c) an
initial tube magnetic strength at a emerge onset depth and d) an
MHD-instability type whose evolution results in a tube ejection
into the solar atmosphere.

In fact, these computations turned out the first basis of the
mechanism for impulsive CME origin as a result of magnetic tubes
(ropes) emerge at high supersonic velocities from the convective
region into the solar atmosphere. They predicted that the
principal difference between impulsive and gradual CME
trajectories should be expected at the initial phase in the solar
atmosphere. A preliminary analysis of experimental data in
\inlinecite{Eselevich2011} has confirmed this conclusion. It
showed that the impulsive CME initial phase is characterized by
the following features:
\begin{enumerate}
\item it starts from the photosphere;
\item the velocity at this phase varies insignificantly with
distance and in time, and may be a few tens through hundreds of km
s$^{-1}$ for various CMEs;
\item the CME angular size at the photosphere level is estimated as
  $d\leq$ (1-3)$^{\circ}$.
\end{enumerate}
In this paper, investigated are the properties of two impulsive
CMEs (one of which being analyzed from the SDO data). We give a
theoretical basis and computation of a possibility to form
impulsive CME as a result of magnetic tubes' ejection from the
convective region into the solar atmosphere at a supersonic
velocity.

\section{Method and data of analysis}
When analyzing, we used the following data:
\begin{itemize}
\item EUV (211\AA) images from AIA/SDO instrument, % of the channel with the 211\AA\ (FeXIV) wavelength,
% \cite{http://www.lmsal.com/get_aia_data/}
temporal resolution being $\approx 12$ sec
(http://www.lmsal.com/get\_aia\_data);
\item EUV (171\AA) images of the full solar disc and corona up to $\approx 1.7R_\odot$ from EUVI
(STEREO/Ahead) with the $\approx 75$ s time resolution
\cite{Howard2008};
\item corona images in the $H_\alpha$ %emission
line of neutral hydrogen obtained by the Digital Prominence
Monitor (DPM) at the Mauna Loa Solar Observatory (MLSO), temporal
resolution being $\sim 3$ min
(http://mlso.hao.ucar.edu/cgi-bin/mlso\_data.cgi).
\end{itemize}
The AIA/SDO brightness data were represented like a difference
brightness $\Delta P = P(t) - P(t_0)$, where $P(t_0)$ is the
undisturbed brightness at $t_0$ prior to the start of the event
under consideration, $P(t)$ is the disturbed brightness at any
moment $t>t_0$. The EUVI/STEREO data were represented like a
running-difference brightness, for which $t$ and $t_0$ are the
moments of the images adjacent in time.
%We studied the CME dynamics using the difference
%brightness images.
We studied the CME dynamics using the difference brightness %images
$\Delta P$ and distributions on the radius $\Delta P(R)$ relative
to the Sun center along the given position angle $PA$ that was
counted off in the images from the North Pole counterclockwise
\cite{Eselevich2008}. The $H_\alpha$ %line
images were represented like brightness isolines
\cite{Eselevich2011}.

\section{Analyzing impulsive CMEs}
 Let us consider two impulsive limb CMEs (longitude of their origin %emergence location
 on the Sun being $\Phi>60^{\circ}$): 25 March 2008 and 13 June 2010. Each of these
CMEs had been already studied earlier
\cite{Gopalswamy2009,Patsourakos2010a,Patsourakos2010b,Temmer2010}.
Therefore, basing on these papers, we will make a special accent
on the initial phase of these events' evolution.

\subsection{25 March 2008 event}
 For this CME, Figure~1A--E
 shows the running-difference images
  %(difference of two images adjacent in time)
  in EUV (from 171\AA\, EUVI/STEREO, Ahead) for six instants
corresponding to this event. At 18:32:15 and 18:34:45 that are not
shown, there is no CME. One manages to reliably record the CME
first occurrence in Figure~1B at 18:38:30: it looks like a
circular cavity with a reduced brightness near the solar surface.
The CME angular size at this moment is $d\approx 3^\circ$ (two
radial straight lines along the CME edges). The position of the
cavity's front boundary (point ``A'' in the $PA = 95^\circ$
direction, median dashed line), coincides approximately with the
vertex (point ``B'') of the bright loop-like structure. The black plus sign %dagger
shows them in Figure~1B (coinciding points ``A'' and ``B''). At
the next moment (Figure~1C) the cavity's front edge ``A'' moves
faster ($V\approx 300$ km~c$^{-1}$), than the loop vertex ``B''
($V\approx 90$ km~c$^{-1}$) and, therefore, further we see them
separately. As we move away from the Sun, a circular frontal
structure (FS), confining the cavity, looks more and more visible
around it. The structure's angular size reaches $d\approx
23^\circ$ when the cavity's leading point ``A'', approximately
coinciding with the frontal structure maximum position, is
recorded at $R_A\approx 1.39R_\odot$. The loop moving more
slowly %(point ``B'')
ceases to be visible inside the cavity by
18:47:15 (Figure~1E).

  \begin{figure} % {1}
  \centerline{\includegraphics[width=0.95\textwidth]
   {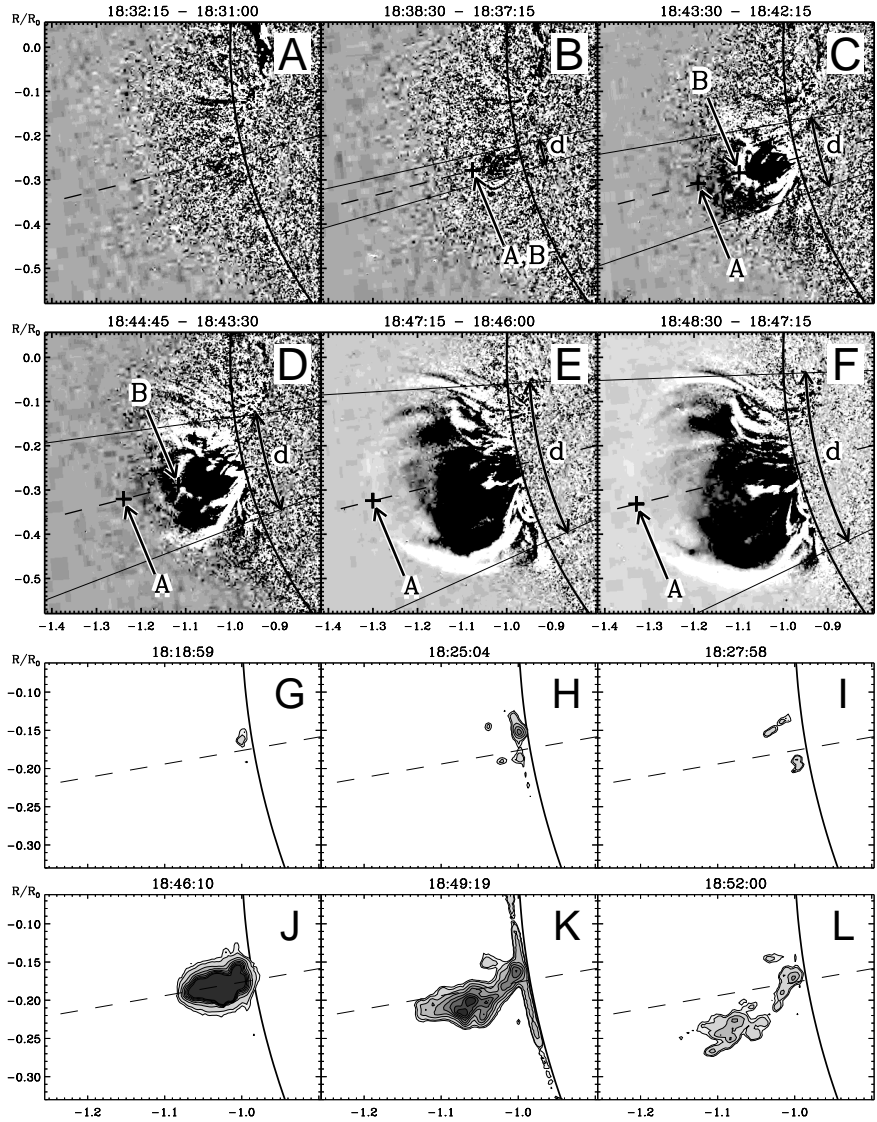}
  }
  \caption{(A-F) The running difference images
  (from 171\AA\ EUVI/STEREO, Ahead data) and (G-L)
  the $H_\alpha$ images in brightness isolines (from DPM/MLSO)
  for different instants corresponding to 25 March 2008 impulsive CME.
    }
  \end{figure}

According to \cite{Patsourakos2010a,Eselevich2011}, the CME
circular frontal structure, when observed in 171\AA\  and in white
light at close instants, approximately coincide in shape and
location, differing only in the frontal structure ring thickness.
It allows to identify as CME frontal structure the bright ring
covering a cavity in Figure~1E.

The leading point ``A'' motion velocity dependencies on distance
$V(R)$ and time $V(t)$ are exhibited in Figure~2A and 2B,
respectively (black circles). One may single out a few
peculiarities characteristic of these dependences.

\begin{enumerate}
\item At the path segment starting with $R\approx 1.04R_\odot$
and up to $R\approx 1.33R_\odot$, the CME cavity leading point
``A'' velocity is almost constant (Figure~2A). (The first-in-time
point in Figure~2A was estimated as per the cavity position at its
first occurrence in the assumption that, at the preceding instant
of measurement, the cavity was at the solar photosphere). In the
$V(t)$ dependence, this segment $t\approx$ 18:37:00 - 18:46:30
corresponds to the ``initial phase''. At distances $R\geq
1.33R_\odot$, the velocity grows fast (Figure~2A) which, in the
$V(t)$ dependence corresponds to the onset of the acceleration
``impulsive phase'', approximately, at 18:46:30 (Figure~2B).

\item The $V(R)$ linear extrapolation up to the photosphere level
(the horizontal dashed line in Figure~2A) provides velocity values
of $V_0\approx 300$ km~s$^{-1}$ close to it.
\end{enumerate}

  \begin{figure} % {2}
  \centerline{\includegraphics[width=0.8\textwidth]
   {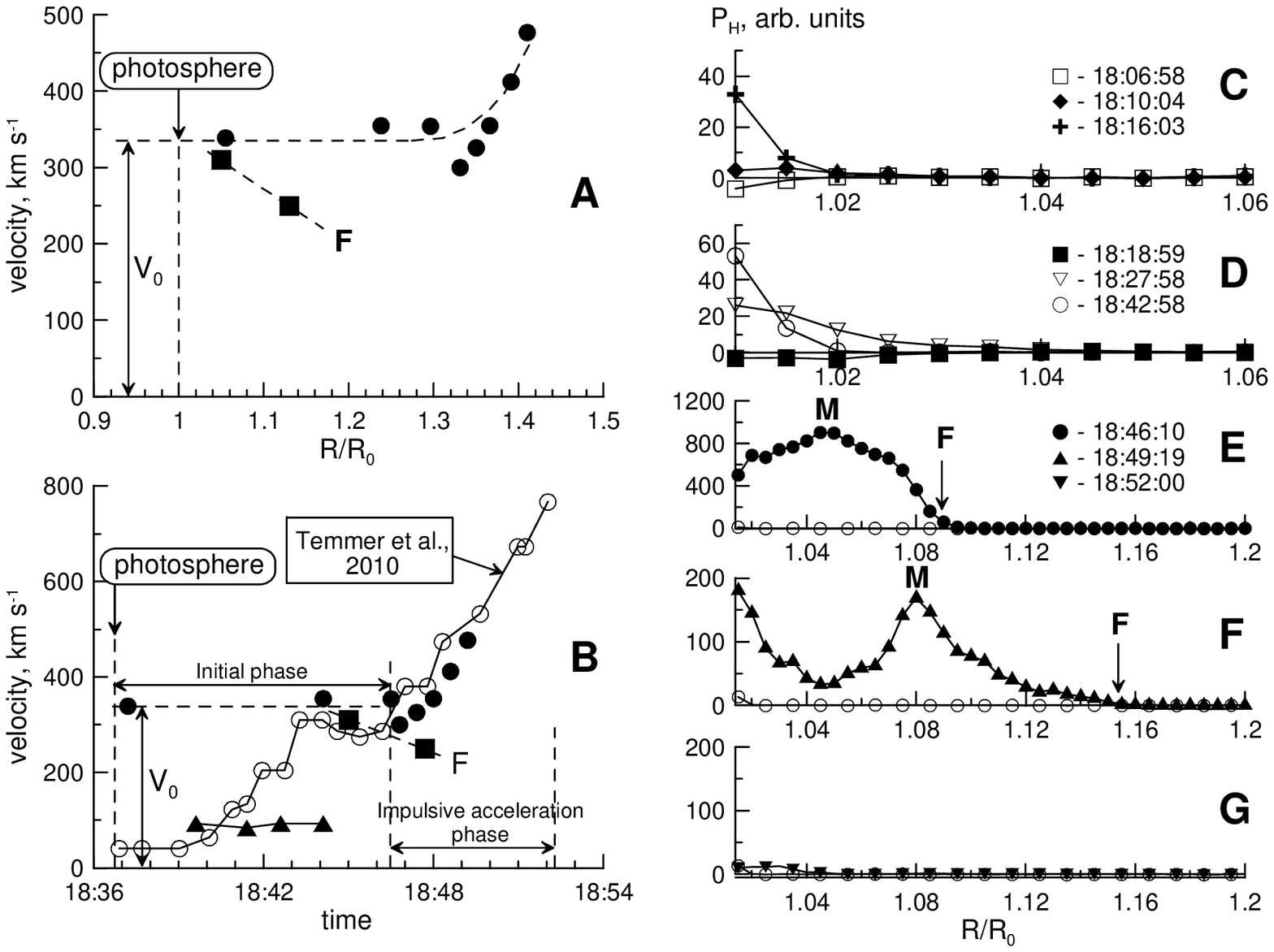}
  }
  \caption{Velocity dependences on distance (A) and time (B)
  of the CME cavity's leading point ``A'' (black circles) and
  leading point ``F'' of the $H_\alpha$ profile $P_H(R)$ along $PA = 100^\circ$
  (black squares);
  (C-G) Radial $H_\alpha$ profiles $P_H(R)$ along $PA = 100^\circ$ for
  consecutive instants on 25 March 2008.
    }
  \end{figure}

The assumed angular size of the CME at the photosphere level is
$d_0\leq 3^\circ$.

 These CME cavity dynamics peculiarities allow
one to interpret it as a manifestation of the magnetic tube (rope)
whose size is under $d_0$ thrown out into the solar atmosphere.
This conclusion agrees with the inferences by
\inlinecite{Patsourakos2010a} that the moving and expanding cavity
(it is termed a bubble there) appear at an instant $t\geq$ 18:35,
but is absent before that, and this is a manifestation of a CME
flux rope, having a smaller size. It allows one to assume that
impulsive CMEs may be thrown out like magnetic ropes from the
convective region at a great velocity essentially exceeding the
speed of sound (speed of sound $V_S\approx$ 6-8 km~s$^{-1}$ at the
photosphere level).

 It is interesting to note that the $V(t)$ curve profile,
starting with 18:43:00 and further, agrees well with the $V(t)$
curve built up from the EUVI (Behind) data in
\inlinecite{Temmer2010} (light circles in Figure~2B). The
difference between black and light circles starts at times under
$t_1\approx$ 18:43:00. Obviously, the reason for this is the fact
that, in the ``Behind'' case at $t <t_1$, the point ``A'' motion
occurs on the solar disc and observing the cavity is impeded
through the presence of bright loop structures whose drift
velocity is actually recorded. In fact, it is the velocity of the
loop vertex ``B'' (black triangles in Figure~2B) thrice as less as
the cavity's front boundary velocity (black circles) that has a
velocity close to the values designated in light circles for
$t <$ 18:42:00. %in our case.

An important additional argument in support of the $V(R)$ linear
approximation up to the solar photosphere for the CME cavity's
front boundary in Figures~2A and 2B is analyzing the kinematics of
a powerful ejection observed in $H_\alpha$ accompanying this CME.
%that we will further be terming Jet.
Let us consider this process
in greater detail. First of all, a few notes on the starting
conditions of this process and terms.

\begin{enumerate}

\item In literature, such
a phenomenon is routinely termed an active prominence or ``surge''
when observed in the $H_{\alpha}$ line and ``jet'' when observed
in EUV and soft X-ray \cite{Priest1982}. To simplify, we will be
terming all the motions in $H_{\alpha}$ {\it jets}.

\item The jet's escape occurred from the NOAA active region 10989
located near the limb (its spot group position is $\approx$
S10~E85).

\item The temperature of quiet, eruptive and active prominence
observed in $H_{\alpha}$ does not exceed $10^4$K \cite{Moore2010}.

\item When observing an active region in
the $H_{\alpha}$ line, even in the absence of CME-like strong
sporadic phenomena or flares, one records motions of a
comparatively cold ($ < 10^4$K) substance at altitudes
significantly smaller than altitudes of quiet prominences
($<15000$ km) \cite{Priest1982} along the magnetic field lines.

\end{enumerate}

 Thus, obviously thereupon that the considered
impulsive CME arose in the active region located near the limb,
its occurrence was preceded by a few very faint ejections of cold
plasma at 50-100 km~s$^{-1}$ within about 30 minutes. In
Figure~1G--I (the $P_H$ brightness isolines in $H_{\alpha}$) they
are seen like weak jets rising and then returning onto the Sun at
18:25:04 and 18:27:58. In Figure~2C,D, we give the examples of
several radial profiles of the brightness in $H_{\alpha}$ towards
$PA =100^\circ$ for weak jets within 18:06:58-18:42:58. Their rise
maximal height for 18:16:03 (pluses) and 18:42:58 (light circles)
did not exceed $h\approx 0.025R_\odot$ above the solar surface.
There was no quasistationary prominence on the limb prior to the
CME emergence. It is seen, for example, in Figure~1G at 18:18:59
and later, i.e., approximately, starting 19 minutes before the CME
first emergence in the cavity's corona.

 The motion of the CME proper is accompanied by a powerful ejection of solar material as
seen in Figure~1J\,--\,L. Let us term it {\it Jet}. Since the
temperature of a weak jet and powerful Jet is circa equal ($\sim
10^4$K) one may estimate the relative mass of the substance
ejected, assuming $M\sim\int P_HdS$, where $S$ is the full area of
the observed jet. It turned out that for the powerful Jet this
value $M_J$ exceeds by almost a factor of 2-3 $M_j$ for weak jets.

The powerful Jet moves, practically, radially until 18:46:10
(Figure~1J), and then swerves, and after 18:49:19 (Figure~1K,L)
returns onto the solar surface. Figure~2E\,--\,G shows radial
profiles of the brightness in $H_{\alpha}$ in consecutive instants
passing through the Jet ($PA = 100^\circ$). These profiles allow
one to trace the Jet dynamics and compare it to the CME dynamics.
A cavity corresponding to the CME is recorded in the solar
atmosphere, for the first time, at 18:38:30 (Figure~1B). There are
no Jet signs in the $H_{\alpha}$ brightness profile corresponding
to the later instant, 18:42:58 (light circles in Figure~2D). This
means that it is the CME that initiates the Jet, but not vice
versa. An unexpectedly huge mass of cold plasma compared with that
of the weak jet also testifies in favor of such a forced mechanism
for the Jet emergence. In the $H_{\alpha}$ brightness profiles
(Figure~2E--G) one can single out the fastest part for the Jet
(F-arrow). A fast reduction of the signal maximum M with time and
then its vanishing in Figure~2E--G reflects the process of the Jet
deviation from the radial direction. (We note that in Figure~2E
the signal is saturated, i.e. the true value of the maximum M may
be essentially greater than the shown value $P_H\approx 900$). The
Jet fastest part velocity is shown by a dashed line with black
small squares in Figure~2A,B. Its initial value near the
photosphere is close to the CME cavity's leading part velocity
(black circles in Figure~2A,B). It confirms the assumption that
the reason for the Jet, most likely, is the CME cavity's escape
from the Sun at a great speed, at least, no lower than the Jet
velocity ($\approx 330$ km~s$^{-1}$).

Let us note that the presence of Jet is not a mandatory signature
during an impulsive CME.

\subsection{13 June 2010 event}
Figure~3A--J shows the difference brightness images (relative to
$t_0$ = 05:32:02) for this CME in EUV (211\AA\ AIA/SDO data) for 9
instants corresponding to this event. The first appearance of the
cavity was observed at 05:32:50 inside a small loop-like structure
(numeral 1 in Figure~3B). Its angular size is $d\approx
1.7^\circ$. This means that the cavity size is slightly smaller.
Such an onset is analogous to the onset of the CME above (25 March
2008). But the further evolution of this event allows one to
reveal a number of interesting peculiarities due to the AIA high
temporal resolution, which enables us to advance in understanding
this phenomenon. Thus, starting with 05:33:50 (Figure~3 does not
show this instant), the formation of four more similar loop-like
Structures starts to become visible ahead of loop-like Structure
1, at different fixed distances, 2, 3, 4 and 5, respectively, in
Figure~3C. In the successive image in Figure~3D, they are seen
more clearly and remain fixed to the instants defined for every
structure.

  \begin{figure} % {3}
  \centerline{\includegraphics[width=0.95\textwidth]
   {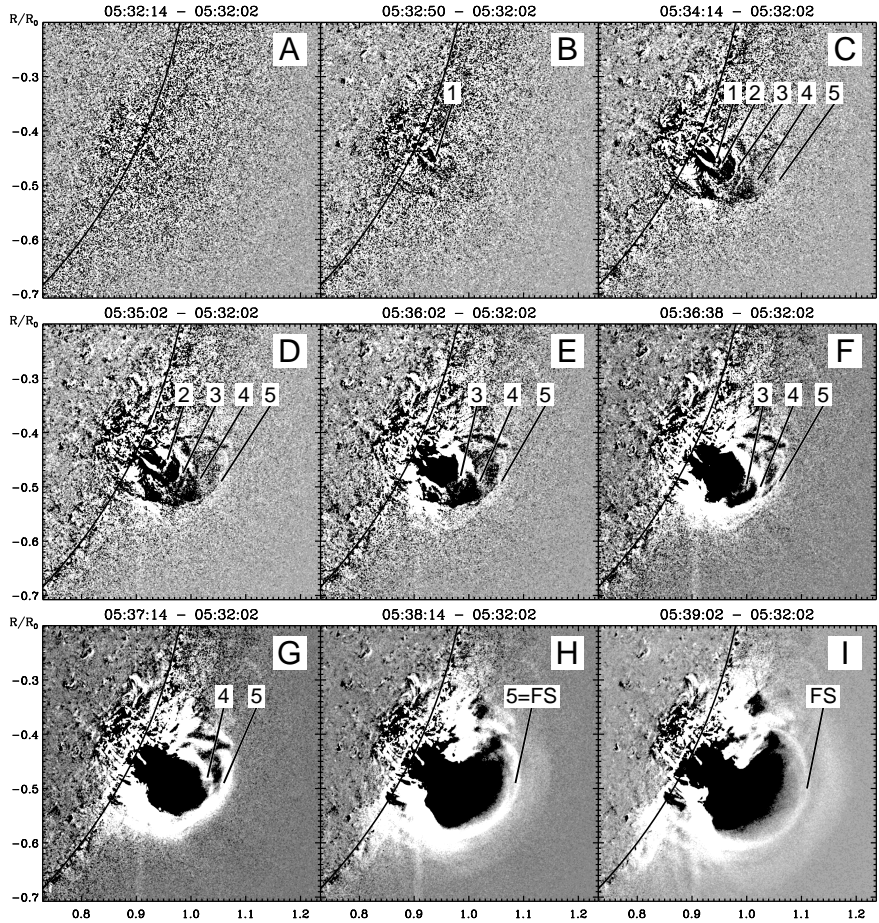}
  }
  \caption{Difference brightness images (relative
  to $t_0$ = 05:32:02) for 9 instants corresponding to
  13 June 2010 impulsive CME (from 211\AA\  AIA/SDO data).
    }
  \end{figure}

The cavity's center in Figure~3B is located at $R_1\approx
1.022R_\odot$ at the first appearance. It is obvious that the
appearance of Structures 2, 3, 4, 5 is associated with the
disturbances that are caused by the cavity appearance in the
corona at $R_1$. Under the impact of the disturbance, the
structures in the difference brightness, invisible previously,
shift and become visible. The first appearance of the farthest
structure 5 is recorded at $t\approx$ 05:33:50 at $R_5\approx
 1.174R_\odot$ (Figure~3 does not show this instant).
 Time dependence of the above-limb height of the five loop-like
structures' vertexes is presented in Figure~4.
 The dashed straight line inclination (left in Figure~4) is determined by the appearance
delay time of every of these structures at different altitudes
$h$. Its crossing with the photosphere provides instant
$t_0\approx$ 05:32:30 of the operation onset for the source of
these disturbances (arrow in Figure~4). At segment $h\approx$
(0-0.2)$R_\odot$, the mean disturbance propagation velocity
$V_{dist}\approx 1600\pm 800$ km~s$^{-1}$, and it is comparable
with the mean Alfven velocity at these distances
\cite{Gopalswamy2009}.

  \begin{figure} % {4}
  \centerline{\includegraphics[width=0.8\textwidth]
   {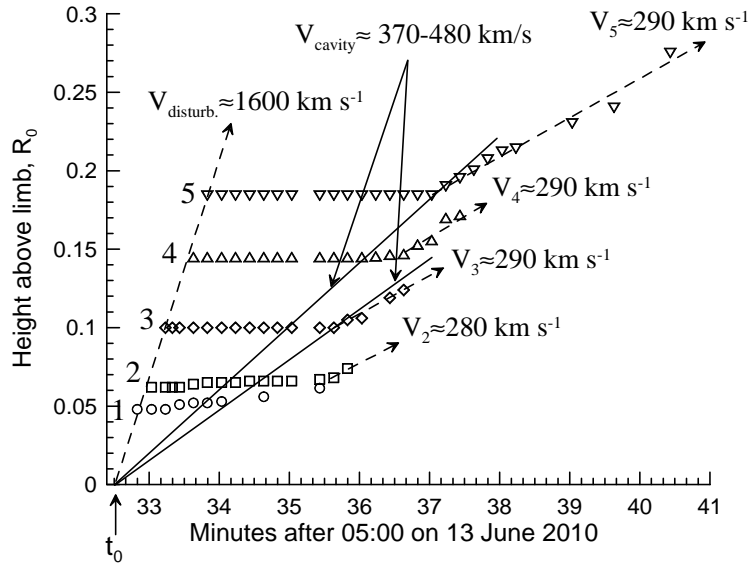}
  }
  \caption{Time dependence of the above-limb height
  of the five loop-like structures' vertexes towards
  $PA$ = (243-244)$^\circ$ from 211\AA\  AIA/SDO data. 13 June 2010 impulsive CME.
    }
  \end{figure}

 The presence of Structures 1-5 formed on the expanding
cavity's motion path is, to a certain extent, an indicator of the
cavity dynamics at these distances. First of all, one may see that
the cavity, moving radially, expands evenly every which way so
that, at all these stages, it is approximately described by a
circle. The visible motion of every structure starts when the
cavity reaches the latter and carries it away. First, the cavity
reaches structure 1 that starts moving at $\approx$ 05:33:20 (in
Figure~4), then is expands and disappears. Structure 2 starts
moving at $\approx$ 05:35:38, then, expanding, disappears.
Structure 3 starts moving at $\approx$ 05:36:02, then, expanding,
disappears. Structure 4 starts moving at $\approx$ 05:36:38,
expanding, it merges with Structure 5 forming the ultimate shape
of the CME frontal structure. The velocities of these structures
reflecting the cavity's propagation velocity make $V\approx$
280-290 km s$^{-1}$ (dashed straight lines in the right part of
Figure~4). It is obvious that the cavity's traveling velocity is
not less than this velocity. The instants of the motion onset (or
the cavity's carrying them away) are most clearly visible for
Structures 3-5. Assuming that the cavity's source and the cavity
proper emerge on the solar surface at $t_0$, we obtain an estimate
of the cavity's mean velocity $V_{cav}\approx$ 370-480 km s$^{-1}$
over the interval $\approx$ 05:32:30-05:37:00 (an inclination of
two bold straight lines in Figure~4). By analogy with the 25 March
2008 'Œ…, it is only natural to assume that a magnetic tube of a
smaller size than the cavity ejected from the convective region in
the solar atmosphere at $V_{cav}\approx$ 370-480 km s$^{-1}$ is
the cavity's source. The obtained value $V_{cav}$ is close to the
CME velocity over the interval 05:37:00-05:38:00 measured in
\inlinecite{Patsourakos2010b} (Figure~4) that made $V\approx$
350-400 km s$^{-1}$.

To summarize this Section, we will note the following. Recent
direct measurements of the polarized signal Doppler shift near the
FeI 5250.217\AA\, line at the IMaX instrument within the Sunrise
stratospheric balloon-borne telescope provided a supersonic value
of magnetic tube rising velocity at the photosphere level $\approx
12$ km s$^{-1}$ \cite{Borrero2010}. Moreover, it was made at the
temporal resolution limit of the instrument. Thus, a possibility
of ejecting magnetic tubes at supersonic velocities into the solar
atmosphere has not only indirect, but also direct experimental
evidence. But, in order to record magnetic tube rising at high
velocities, it is necessary to increase the temporal resolution of
present-day instruments.

\section{Discussing the mechanism for origin of impulsive CMEs}

From the stated above it follows that impulsive CMEs may gain
initial velocity in the convective region where the medium plasma
is ideal and optically thick. In this region, strong frozen
magnetic field represents an ensemble of magnetic flux ropes
(magnetic tubes) \cite{Stenflo1973}, and the plasma motion is well
described in the ideal MHD approximation. It imposes some
characteristic restrictions on the problem: the plasma motion is
self-consistent and this does not allow one to simply bring in
additional power sources or field twists like, for example, it is
the case when considering magnetic ropes in the corona
\cite{Chen1996}. All the sources require a complete substantiation
of their nature. Under these circumstances, the problem to search
for a mechanism for forming impulsive CMEs is put as an as a
problem with initial conditions, after which the tube evolution is
described self-consistently by a set of MHD equations. To describe
the dynamics of a separate magnetic tube, we use an assumption
that the tube parameters across the cut are circa homogenous. This
assumption is well-substantiated (droppable values have the
infinitesimal order of $\sim 10^{-2}$\,--\,$10^{-6}$).

When transiting from of the MHD three-dimensional system to the
thin magnetic tube approximation \cite{Romanov1993a}, one takes
into account the cross balance of internal ({\it i}) and external
({\it e}) pressures
%inside ({\it i}) and outside ({\it e}) of the tube:
\begin{equation}
\label{balance} p_i+\frac{H^2}{8\pi} = p_e,
\end{equation}
after which an MHD motion equation
\begin{equation}
\label{motion} \rho \left[\frac{\partial \vec v}{\partial t} +
(\vec{v},\nabla) \vec v\right] = - \nabla \left[p +
\frac{H^2}{8\pi}\right] + \frac{1}{4\pi}(\vec{H},\nabla)\vec{H} +
\rho \vec{g}
\end{equation}
assumes the form
\begin{equation}
\label{motion2} \rho_i \frac{d\vec{v}}{d t} = - \nabla p_e +
\rho_i \vec{g} + \frac{1}{4\pi}(\vec{H},\nabla)\vec{H}.
\end{equation}
The main forces are the pressure gradient, gravity and tensile
force of the magnetic field. In a hydrostatic medium ($ \nabla p_e
= \rho_e g $) the first two unite into the buoyancy force $
(\rho_i - \rho_e) g $.

Papers \cite{Romanov1993a,Fan1994} considered the dynamics of a
similar magnetic tube rising in detail. Below, we present the
principal results most relevant in terms of studying the initial
stage of impulsive CME formation.

\begin{enumerate}
\item  At lateral compression, a part of the external pressure is
counterbalanced by the magnetic field $p_{mag} = H^2 / (8\pi)
\propto \rho^2$, which changes the effective adiabatic index of
the magnetized plasma and results in that the convective
instability increment is nonzero.
\item  The ``slow wave'' instability
wave (the Parker instability) is the key instability. Its
mechanism is as follows: at a form slight disturbance, the tube
plasma flows downward from the rising part. If the disturbance
wave length is great enough, the field tensile force is not
capable of compensating the buoyancy force upward on the radius,
and the total pressure gradient does not hinder with the plasma
flow along the tube. As a result, the loop vertex continues rising
with an increasing velocity.
\item The input parameters of the
problem are only the tube's initial state near the convective
region bottom and distribution of the outer medium parameters; the
system of equations has been obtained directly from MHD and is
self-consistent.
\end{enumerate}
 The latter condition is a strong point of the
model since without bringing in model sources and by only
selecting the initial state we reproduce simultaneously a wide
spectrum of the observed peculiarities: range of latitudes for
sunspot emergence; sunspot inclination angle towards the equator
and its dependence on the field strength and position; asymmetry
between the western and eastern sunspots, etc. All this justifies
the quality of the model.

 Figure~5 presents the result of modeling
a rising of a magnetic tube from the rest state
\cite{Romanov2010}. Initially, the tube rests horizontally in the
equatorial plane. Affected by the Parker instability, an arc whose
vertex rises upward is formed and punches the photosphere.
Although above the photosphere the used equation of energy is
already inapplicable (radiant heat exchange becomes volumetric),
the accumulated impulse changes mainly under the influence of the
field tensile force and the buoyancy force so that the results are
a good order-of-magnitude estimate. The tube reaches the
photosphere  fast, with a peak velocity of the order of 100
km~s$^{-1}$ (Figure~5).

  \begin{figure} % {5}
  \centerline{\includegraphics[width=0.7\textwidth]
   {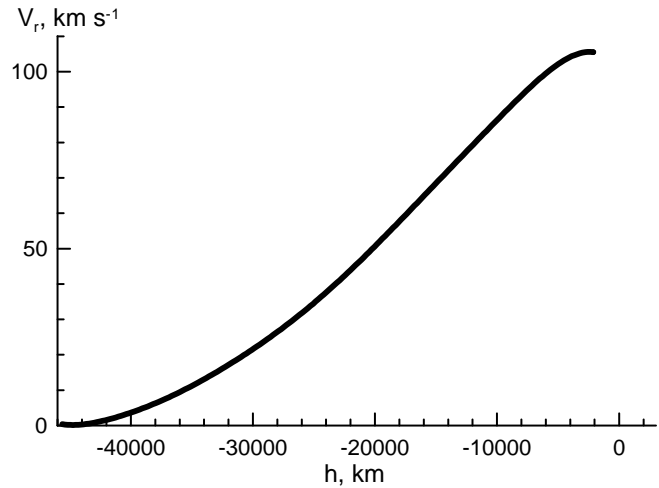}
  }
  \caption{Radial velocity $V_r$ (computed in MHD-approximation)
  of the magnetic tube rising from the rest state depending on the ``depth'' $h$
  relative to the photosphere level.
    }
  \end{figure}

\section{Conclusions}
\begin{enumerate}
\item  We have confirmed the key difference
established earlier in \inlinecite{Eselevich2011} between the
impulsive and gradual CME trajectory ``initial phase''.
\item We have arrived at the conclusion that forming impulsive CMEs
starts under the solar photosphere and may be associated with
supersonic emergence  of magnetic tubes (ropes) from the
convective region. At the photosphere level, the radial velocity
of such tubes may reach up to hundreds of km~s$^{-1}$, and their
angular size may be $d\approx$ (1-3)$^\circ$.
\item  A probable reason for supersonic rise of magnetic tubes (ropes) from the convective
region is the ``slow wave'' instability (the Parker instability).
\end{enumerate}

Acknowledgements. We thank the teams controlling all the
instruments whose data have been used in this study for their
efforts and open data policies: the ESA and NASA STEREO/SECCHI and
SDO/AIA telescopes. Courtesy of the Mauna Loa Solar Observatory,
operated by the High Altitude Observatory, as part of the National
Center for Atmospheric Research (NCAR). NCAR is supported by the
National Science Foundation. The research was supported by the
Russian Foundation of Basic Research Grants No. 09-02-00165a,
No.10-02-00607a. We thank Yuri Kaplunenko for the help in
translation into the English.

%Courtesy of STEREO/SECCHI team, NASA/SDO and the AIA science team,
%the Mauna Loa Solar Observatory, operated by the High Altitude
%Observatory, as part of the National Center for Atmospheric
%Research (NCAR). NCAR is supported by the National Science
%Foundation.

\end{article}
\end{document}